\documentclass[prl,twocolumn,showpacs,preprintnumbers,amsmath,amssymb]{revtex4}

\usepackage{graphicx}
\usepackage{dcolumn}
\usepackage{bm}
\usepackage{graphics}
\input{epsf}

\newcommand{\beq}{\begin{equation}}
\newcommand{\eeq}{\end{equation}}
\def\la{\hbox{\raise.35ex\rlap{$<$}\lower.6ex\hbox{$\sim$}\ }}
\def\ga{\hbox{\raise.35ex\rlap{$>$}\lower.6ex\hbox{$\sim$}\ }}

\def\beq{\begin{equation}}
\def\eeq{\end{equation}}
\def\beqa{\begin{eqnarray}}
\def\eeqa{\end{eqnarray}}
\def\bseq{\begin{subequations}}
\def\eseq{\end{subequations}}

\def\order#1{{\cal O}\left({#1}\right)}
\newcommand{\sfrac}[2]{\mbox{$\frac{#1}{#2}$}}
\begin{document}
\preprint{APS/123-QED}
\date{\today}

\title{A weakly nonlinear analysis of the magnetorotational instability
in a model channel flow}
\author{O.M. Umurhan$^{1,2,4}$}
\author{K. Menou$^{3}$}
\author{O. Regev$^{1,3}$}
\affiliation{$^{1}$Department of Physics, Technion-Israel Institute of
Technology,Haifa, Israel \\
$^{2}$Department of Geophysics and Planetary Sciences, Tel--Aviv
University, Tel-Aviv, Israel \\
$^{3}$ Department of Astronomy, Columbia University, New York, NY
10027, USA \\
$^{4}$Department of Astronomy, City College San Francisco, SF, CA,
94112 USA}

\begin{abstract}
 We show by means of a
perturbative weakly nonlinear analysis that the axisymmetric magnetorotational
instability (MRI)
of a viscous, resistive, incompressible rotating shear flow in a
thin channel gives rise to a real Ginzburg-Landau equation
for the disturbance amplitude. For small magnetic Prandtl number
(${\cal P}_{\rm m}$), the saturation amplitude is $\propto
\sqrt{{\cal P}_{\rm m}}$ and the resulting momentum transport
scales as ${\cal R}^{-1}$, where $\cal R$ is the {\em
hydrodynamic} Reynolds number. Simplifying assumptions, such as
linear shear base flow, mathematically expedient boundary
conditions and continuous spectrum of the vertical linear modes,
are used to facilitate this analysis. The asymptotic results are
shown to comply with numerical calculations using a spectral code.
They suggest that the transport due to the nonlinearly developed
MRI may be very small in experimental setups with ${\cal P}_{\rm
m} \ll 1$.

\end{abstract}

\pacs{Valid PACS appear here}
\maketitle
The magnetorotational instability (MRI) has acquired
growing theoretical and experimental interest in recent years,
moving beyond just
the astrophysical community. The linear MRI has been known for almost 50 years
\cite{veli,chandra1}: Rayleigh stable rotating Couette flows
of conducting fluids are
destabilized in the presence of a vertical magnetic field
 if $d\Omega^2/dr<0$
(angular velocity decreasing outward). However, the MRI only
acquired importance to astrophysics after the influential work of
Balbus \& Hawley \cite{baha1}, who demonstrated its viability to
accretion disks (for any sufficiently weak field) with linear
analysis supplemented by nonlinear numerical simulations. Enhanced
transport (conceivably turbulent) is necessary for accretion to
proceed in disks found in a variety of astrophysical settings -
from protostars to active galactic nuclei. The difficulty in
identifying sources of such enhanced transport has been a
limitation to proper theoretical understanding and modeling of
these objects. Thus the MRI has been widely accepted as an
attractive solution for enhanced transport (see the comprehensive
reviews \cite{rev1,rev2}) even though some questions on the nature
of MRI-driven turbulence remain \cite{brand}. Because of the MRI's
importance and some of its outstanding unresolved issues, several
groups have recently undertaken projects to investigate the
instability under laboratory conditions \cite{jgk,colgate,sisan}.
Additionally, numerical simulations specially designed for
experimental setups have been conducted.

In this Letter we perform a weakly nonlinear analysis of the MRI
near threshold
for a rotating flow in channel geometry, subject to
idealized but mathematically expedient boundary conditions.
This kind of approach is important because the
viability of this linear instability as the driver of
turbulence and angular-momentum transport
relies on understanding its
nonlinear development and saturation.
By complementing the above mentioned simulations,
analytical methods remain useful to
gain further physical insight. To
facilitate an analytical approach we make a number of simplifying
assumptions so as to make the system amenable to well-known
methods \cite{BO,crossandI,manbook,mybook} for the
derivation of nonlinear envelope equations
(i.e. with an amplitude weakly dependent on
time \emph{and} space).  Additionally we assume a narrow channel
geometry, i.e. the small-gap limit of Taylor-Couette flow.

The hydromagnetic equations in cylindrical coordinates
\cite{chandra} are applied to the neighborhood of a representative
point in the system, using the shearing box (SB) approximation
\cite{goldreich65}. The SB has been employed in numerous
(analytical and numerical) studies of the MRI in accretion disks
and are also appropriate for this model problem. Within the
framework of the {\em small} shearing box (SSB)\cite{SSB}, the
base flow is steady and incompressible with constant pressure. The
velocity ${\bf V}=U(x){\bf {\hat y}} $ has a linear shear profile
$U(x)=-qU_0 x$, representing an axisymmetric flow about a point
$r_0$, that rotates with a rate $\Omega_0$ defined from the
differential rotation law $\Omega(r) \propto \Omega_0
(r/r_0)^{-q}$. We restrict ourselves to a constant initial
magnetic field ${\bf B} = B_0 {\bf \hat z}$. Cartesian coordinates
$x,y,z$ are used to represent the radial (shear-wise), azimuthal
(stream-wise) and vertical directions, respectively. The base flow
is axisymmetricaly disturbed by perturbations of the
velocity, ${\bf u}=(u_x,u_y,u_z)$, magnetic field, ${\bf b}=
(b_x,b_y,b_z)$, and total pressure, $\varpi$. This results, after
non-dimensionalization, in the following set of {\em non-linear}
equations:
\begin{subequations}
\label{full_equations}
\beqa
& & \frac{d{\bf u}}{dt}-
2{ { \bf{ \hat z}}\times{\bf u}}
-{q   u_x {\bf{\hat y}}}
- {\cal C}\,\hat{\mathbb B}\,{\bf b} = -\nabla\varpi
+\frac{1}{{\cal R}}\nabla^2 {\bf u} \ \ \ \ \
\label{eq1}
\\
&& \frac{d{\bf b}}{dt} -
\hat{\mathbb B}\,{\bf u}
+
{q  b_x {\bf{\hat y}}} =
\frac{1}{{\cal R}_{{\rm m}}}\nabla^2 {\bf b} \\
&&\nabla\cdot{\bf u} = 0~~~~  {\rm and}  ~~~~ \nabla\cdot{\bf b} = 0,
\label{eq3}
\eeqa
\end{subequations}
where $\hat{\mathbb B}\equiv({\bf b}\cdot\nabla + B_0\partial_z)$.
Lengths were scaled by $\tilde L$ (the channel width scale), time
by $\tilde \Omega_0^{-1} \equiv \Omega_0^{-1}$ and the magnetic
field by $\tilde B_0 \equiv B_0$. The quantities marked by tilde
are thus dimensional. With this scaling, $B_0 =1$ in (1)
but we retain it  for later convenience. The non-dimensional
parameter $ {\cal C} \equiv \tilde B_0^2/(4\pi\tilde \rho_0 \tilde
\Omega^2 \tilde L^2) \equiv \tilde V_A^2/\tilde U^2,
\label{alfven_number} $ is the {\em Cowling number}, measuring the
dynamical importance of the magnetic field. In addition the {\em
Reynolds number}, $ {\cal R} \equiv\tilde\Omega_0 \tilde
L^2/\tilde \nu, $ and the {\em magnetic Reynolds number}, $ {\cal
R}_{\rm m} \equiv \tilde\Omega_0 \tilde L^2/\tilde \eta, $ are
used, where $\tilde \nu$ and $\tilde \eta$ are the microscopic
kinematic viscosity and magnetic resistivity of the fluid,
respectively. We introduce two additional non-dimensional
parameters, which figure in what follows: the \emph{Elsasser
number}, $ {\cal S} \equiv {\cal R}_{\rm m} {\cal C} = \tilde
B_0^2/(4\pi\tilde\rho_0 \tilde \Omega_0\tilde\eta), $ and the {\em
magnetic Prandtl number}, ${\cal P}_{\rm m} \equiv {\cal R}_{\rm
m}/{\cal R}$.
\par
Linearization of (1), for perturbations of the
form $\propto e^{st + ik_x x + ik_z z}$, gives rise to the dispersion relation,
\[
a_0s^4 + a_1s^3 + a_2s^2 + a_3s + a_4 = 0,
\]
where all the $a_i = a_i(k_x,k_z,{\cal P}_{\rm m},{\cal S},{\cal C},q)$
but we shall write
explicitly only $a_4$:
\beq
 \frac{{\cal C}}{{\cal S}^4}\Bigl[
k_{_T}^2{\cal C}({\cal C}k_{_T}^4 {\cal P}_{\rm m} + k_z^2 {\cal S}^2)^2
+\kappa^2 {\cal S}^2{\cal C} k_{_T}^4 k_z^2 -2 q   {\cal S}^4k_z^4
 \Bigr],
 \label{a_coefficients}
\eeq
where the notation $\kappa^2 \equiv
2(2-q), \ k_{_T}^2 \equiv k_x^2 + k_z^2$
is introduced.

For given values of the parameters (we denote them collectively as
$\Pi$) there will be four distinct modes. The linear theory in
various limits for this problem has been discussed in numerous
publications (see \cite{rev1,rev2}) and we shall not elaborate
upon them any further. Rather, we focus on situations where the
most unstable mode (of the four) is marginal (critical) for some
$k_z$ at a given value of $k_x = K$. We fix $K$ in order to
focus on the marginal vertical dynamics within a channel (see below).

Marginality to the MRI ($s=0$) implies the vanishing
of the real coefficient $a_4$ (as expressed above)
and its derivative with respect to $k_z$ at some $k_z=Q$:
\beq
a_4(k_z=Q; K, \Pi)=0, \ \  \frac {\partial a_4}{\partial k_z}(k_z = Q; K, \Pi)=0.
\eeq

The second condition and (\ref{a_coefficients}) yield
\beqa
& & Q\Bigl[Q^2{\cal S}^4({\cal C}Q^2 - 4q )
+(K^2 + Q^2)2Q^2{\cal C}{\cal S}^2({\cal S}^2+ \kappa^2) \nonumber \\
& & +(K^2 + Q^2)^2{\cal C}{\cal S}^2(6{\cal C}{\cal P_{\rm
m}}Q^2 + \kappa^2)
+2(K^2 + Q^2)^3{\cal P_{\rm m}}{\cal C}^2{\cal S}^2 \nonumber \\
& & +5(K^2 + Q^2)^4{\cal P_{\rm m}}^2{\cal C}^3\Bigl] =0. \label{da4dkz}
\eeqa
This equation, together with $a_4(k_z=Q, K; \Pi)=0$, can be solved
for ${\cal S}$
and $Q$.
The general expressions for ${\cal S}(K,{\cal P}_{\rm m},{\cal C},q)$
and $Q(K,{\cal P}_{\rm m},{\cal C},q)$ are lengthy
but their asymptotic forms,
for ${\cal P}_{\rm m} \ll 1$, to leading order,
\beq
{\cal S} = \frac{\sqrt{16 \ {\cal C}q(2-q)} K}{(2q-{\cal C}K^2)},\ \
 Q^2
= K^2\frac{2q-{\cal C}K^2}{2q+{\cal C}K^2}.
 \label{def_S}
\eeq
If ${\cal C} K^2 > 2q$ the solutions are not physically meaningful,
while the case
${\cal C} K^2 = 2q$ corresponds to the ideal MRI limit.
\par
We consider the properties of the MRI within the confines
of an idealized model channel geometry with walls located at
$x=0,\pi/K$. We allow all quantities to be vertically periodic on
a scale $L_z$ commensurate with integer multiples of $2\pi/Q$. As
long as $L_z \gg 1/Q$, the limit of a vertically extended
system (and thus a continuous spectrum of vertical modes) may be
affected.
\par
We follow the fluid into instability by tuning the vertical
background field away from the steady state, i.e.
$ B_0
\rightarrow 1 - \epsilon^2\tilde\lambda, $
where $\epsilon \ll 1$ and $\tilde \lambda$ is an $\order 1$ control parameter.
This means that we are now in a position to apply procedures of
singular perturbation theory to this problem,
by employing the multiple-scale (in $z$ and $t$) method (e.g,
\cite{BO}). It facilitates,  by imposing suitable solvability
conditions at each expansion order of the calculation in order
to prevent a breakdown in the solutions, a derivation of an
envelope equation for the unstable mode.
Thus for any dependent fluid quantity $F(x,z,t)$ we assume
\beq
F(x,z,t) = \sum_{i=1}^\infty \epsilon^iF_i(x,z,t).
\label{expansion_scheme}
\eeq
The fact that the $x$ and $z$ components
of the velocity and magnetic field perturbations can be derived from
a streamfunction, $\Psi(x,z)$ and magnetic flux function $\Phi(x,z)$ reduces
the number of relevant dependent variables $F$ to four ($u_y$ and $b_y$
are the additional two). These four dependent variables will also be used
in the spectral numerical calculation (see below).

For the lowest $\epsilon$ order of the equations, resulting from substituting
the expansions into the original PDE and collecting same order terms, we make
the Ansatz
$
F_1(x, z,t) = \hat F_1 \tilde A(\epsilon z,\epsilon^2 t)e^{iQ z}\sin K x
+ {\rm c.c.},
$
where $\hat F_1$ is a constant (according to the variable in question), and
where the envelope function $\tilde A$
(an arbitrary constant amplitude in linear theory)
is now allowed to have weak space (on scale $Z\equiv \epsilon z$)
and time (on scale $T \equiv \epsilon^2 t$) dependencies.
Because
this system is tenth order in $x$-derivatives,
a sufficient number of conditions must be specified at the edges.
The $\sin K x$ functional dependence of the
Ansatz satisfies the ten boundary conditions
\beq
u_x=u_y=b_x=b_y=\partial_x u_z = 0, \label{boundary_conditions}
\eeq at $x=0,\pi/K$.
These conditions are also subsequently
satisfied to all orders of the expansion
procedure because of the symmetry
property of $\sin K x$ under the nonlinear operations of
(\ref{full_equations}).
These specifications are a mixture of kinematic conditions
on the edges: no-slip conditions on $u_x$ and $u_y$ with
free-slip conditions on $u_z$. The conditions are
similarly mixed for the magnetic field: conducting conditions on
$b_x$ and insulating conditions on $b_y$. Although these
conditions are idealized,
they are mathematically expedient, in that the coefficients of the resulting
envelope equation may be expressed analytically in the limit
${\cal P}_{\rm m}\ll 1$, and they serve to capture the salient features of the
resulting dynamics (see below).

The end result of the asymptotic procedure procedure is the well-known
real Ginzburg-Landau Equation (GLE) which, for
${\cal P}_{\rm m} \ll 1$, is
\beq
\partial_T A = \lambda A
- \frac{1}{{\cal P}_{\rm m}{\cal C}} A|A|^2 + D \partial_{_Z}^2 A .
\label{landau_eqn} \eeq Here  $A \equiv {\sqrt \xi} \tilde A $,
$\lambda \equiv \zeta \tilde\lambda $, and the coefficients, for a
value of $q=3/2$ (corresponding to a Keplerian shear flow) are
\begin{subequations}
\beqa \xi &=&
\frac{3}{4}\cdot\frac{
5{\cal S}^4 - 18{\cal S}^2 - 32
+ 2({\cal S}^2 + 16)\sqrt{{\cal S}^2 + 1}}
{{\cal S}({\cal S}^2 +1)(4\sqrt{{\cal S}^2 +1}-3)}
, \ \ \ \\
D &=& 6\frac{\left({\cal S}^2 +2 - 2\sqrt{{\cal S}^2+1}\right)({\cal
S}^2 + 1)} {{\cal S}^3(4\sqrt{{\cal S}^2+1} - 3)},\\
\zeta &=& \frac{3{\cal S}\sqrt{{\cal S}^2 +1} - 6{\cal S}}
{4{\cal S}^2 +1 +\sqrt{{\cal S}^2+1}},
\label{landau_coefficients}
\eeqa
\end{subequations}
where ${\cal S}(K)$ is as given in (\ref{def_S}).
The expressions for a general $q$ are very long and involved.
For ${\cal S} \gg 1$, (i.e. as one approaches the ideal MRI
limit), these
simplify to $ \xi =15/16, \ \zeta =3/4, \
D = 3/2$; and in general they remain $\order 1$ quantities for all
reasonable values of ${\cal S}$.


Before proceeding any further, it is important to note that
equations (1) admit an integral relationship for the total
disturbance energy. Application of the
boundary conditions (\ref{boundary_conditions})
yields a hydromagnetic extension of the
Reynolds-Orr equation \cite{sh}. Here it takes on the form:
\beqa \frac{d{\rm E}_{\rm V}}{dt} &=& -\frac{ dU}{dx} \int (u_x u_y -{\cal C}
b_x b_y) dx dz \nonumber \\
 & & -\int{\left(\frac{1}{{\cal R}}|\nabla {\bf u}|^2
+\frac{1}{\cal R_{{\rm m}}}{\cal C} |\nabla {\bf b}|^2\right)
 dx dz}, \ \ \ \ \
\label{energy_theorem} \eeqa where $|\nabla {\bf u}|^2 \equiv
\sum_\mu |\nabla u_\mu|^2$ for $\mu=x,y,z$ and similarly
for ${\bf b}$. ${\rm E}_{\rm V}$ is the total energy, per unit length in the
azimuthal direction, of the disturbances in the
domain, ${\rm E}_{\rm V} \equiv \sfrac{1}{2}\int{\left({\bf u}^2 + {\cal C}{\bf
b}^2\right)dxdz}$, and the first integral (multiplied by $dU/dx$)
represents the energy fed into the
hydrodynamical and magnetic disturbances by the background shear.
The integrand ${\rm T} \equiv u_x u_y -{\cal C} b_x b_y$ is the relevant
component of the total stress in the disturbance.
Integrating it along the wall and averaging radially
gives an estimate of the average angular momentum transport rate
inside the channel, i.e. ${\rm{\dot J}}\equiv
(K/\pi)\int_{-L_z/2}^{L_z/2}\int_{0}^{\pi/K}{{\rm T}dxdz }$.
\par
We can use our asymptotic solutions to express ${\rm E_{\rm V}}$ and $\dot{\rm J}$ in
terms of the envelope function $A(Z,T)$.
To the first few leading orders in $\epsilon$ (recalling
that ${\cal P}_{\rm m} \ll 1$) we find that
\[
{\rm E}_{\rm V} = \epsilon^2 \alpha({\cal S}) {\cal C}^2 A^2 +
\epsilon^4 {\cal C}^3 \left[ \frac{\beta_{2}({\cal S})}{{\cal P}_{\rm m}^2}
+\frac{\beta_1({\cal S})}{{\cal P}_{\rm m}} \right] A^4  + \cdots.
\]
$\alpha$ and $\beta_i$ are functions whose value is of $\order 1$ for all
reasonable values of ${\cal S}(K,q=3/2)$. The two bracketed terms
arise from the second order
(in $\epsilon$) azimuthal velocity disturbances. Contributions to the
total angular momentum transport ($\dot {\rm J}= \dot {\rm J}_{\rm H} +
\dot {\rm J}_{\rm B}$)
due to the hydrodynamic
($\dot {\rm J}_{\rm H}$) and magnetic
correlations ($\dot {\rm J}_{\rm B}$) are,
to leading order,
\begin{subequations}
\beqa
\dot {\rm J}_{\rm H} &=& \frac{9\epsilon^2}{{\cal S}}\left(
\frac{2+{\cal S}^2 - 2\sqrt{1+{\cal S}^2}}{1+{\cal S}^2}
\right)A^2
+\order{\epsilon^3,\epsilon^2{\cal P}_{\rm m}}, \ \ \ \ \
\label{angular_momentum_flux}
\\
\dot {\rm J}_{\rm B} &=& \frac{3\epsilon^2}{{\cal S}}\left(
1
-\frac{1}{\sqrt{1+{\cal S}^2}}\right)A^2
+\order{\epsilon^3,\epsilon^2{\cal P}_{\rm m}}.
\eeqa
\end{subequations}

\par
The envelope $A$ is found by solving the  GLE, which is a
well-studied system (see, e.g., \cite{mybook} for a summary and references).
It has three steady uniform solutions in one spatial dimension (here, the vertical):
an unstable state $A = 0$, and two stable states $A=\pm A_s$,
where $ A_s^2 = |\zeta({\cal S})|{\cal P}_{\rm m}{\cal C}$ (note that
$\zeta({\cal S})$ is an $\order{1}$ quantity).
$A_s$ is also the saturation amplitude,
because the system will develop towards it \cite{mybook}.

Setting $A \rightarrow A_s$ in
(\ref{angular_momentum_flux}-b) and in the expression for ${\rm E_V}$, followed by some algebra,
reveals that the angular momentum flux in the saturated state is, to leading order in
${\cal P}_{\rm m}$ and $\epsilon$,
$\dot{\rm J} =  \epsilon^2 |\zeta({\cal S})|
\gamma({\cal S}){\cal P}_{\rm m} {\cal C}{\cal S}^{-1}$,
while in the expression for the energy one term is independent
of ${\cal P}_{\rm m}$,
${\rm E_V}=\epsilon^4 {\cal C}^3 \zeta^2({\cal S}) \beta_2({\cal S}) +\order{{\cal P}_{\rm m}}$.

The key results of this idealized analysis are thus,
\beq
A_s \sim \sqrt{{\cal P}_{\rm m}{\cal C}} \longrightarrow
{\rm E_V}\sim E_0,~~~~  \dot{\rm J}\sim {\cal R}^{-1},
\eeq
for ${\cal P}_{\rm m}\ll 1$,
where $E_0$ is a constant, independent of ${\cal P}_{\rm m}$.

We have also performed numerical calculations, using a
2-D spectral code to solve the original
nonlinear equations (1), in the streamfunction and magnetic flux
function formulation, near MRI threshold.
The asymptotic theory reflects the trends seen in these
simulations.
The code implements a Fourier-Galerkin expansion
in ${\rm 64}\times{\rm 64}$ modes in each of
the four independent physical variables, i.e.
${\bf F}= \sum_{n,m}{\bf F}_{\rm nm}(t)
\sin K_n x e^{iQ_mz} + {\rm c.c.}$,
where ${\bf F} = (\psi,u_y,\Phi,b_y)^{{\rm T}}$
and where
${\bf F}_{\rm nm}(t)$ is the time-dependent amplitude
of the particular Fourier-Galerkin mode in question
(denoted by the indices $n,m$).

\begin{figure}
\begin{center}
\leavevmode \epsfysize=4.0cm
\epsfbox{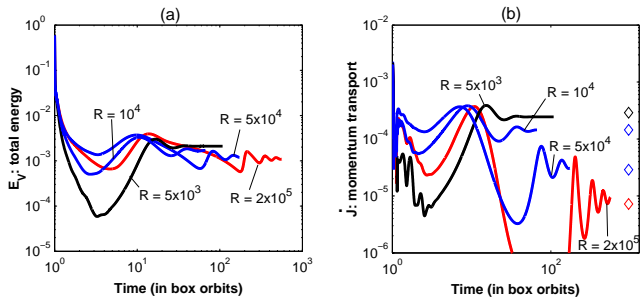}
\end{center}
\caption{{\small
${\rm E_V}$ (panel a) and $\dot {\rm J}$ (panel b) as a function of time
from numerical calculation. The black line is for ${\cal R} = 5\times10^3$,
the red one for ${\cal R} = 2\times10^5$.  The intermediate
values (in blue) are successively for ${\cal R} = 10^4,5\times10^4$.
The diamonds in panel (b) show the scaling predicted by our asymptotic
analysis, which
also predicts a constant final value of the disturbance energy, as apparent in panel (a).
One box orbit $=2\pi$ in nondimensionalized units.
}}
\label{numerics_plots}
\end{figure}

We typically start with
white noise initial conditions
on ${\bf F}_{\rm nm}(0)$
 at a level of 0.1
the energy of the background shear.
Because of space limitations we display
only a representative plot of runs made with
${\cal S} = 5.0$, ${\cal C} = 0.05$, (i.e. ${\cal R}_{\rm m}$ fixed),
$q=3/2$
and $\epsilon^2 = 0.2$ for a few successively increasing
values of ${\cal R}$ (see Figure \ref{numerics_plots}).
Note that with this value of $\epsilon$
the fastest growing linear mode has a growth rate of $\sim 0.065$
in these units, and thus fully developed ideal MRI can not be
expected.
It is apparent that ${\rm E}_{\rm V}$ saturates at a constant independent of
${\cal R}$ (for large enough $\cal R$, i.e. small enough ${\cal P}_{\rm m}$),
while $\dot{\rm J}$ saturates at values that scale as ${\cal R}^{-1}$,
as expected from our asymptotic analysis.

The numerical and asymptotic solutions developed here
show that
in the saturated state the second order
azimuthal velocity perturbation
becomes dominant over all other quantities for ${\cal P}_{\rm m}
\ll 1$: it has the spatial form $\sin 2K x$, independent of $z$.
It arises from the $\order{\epsilon^2}$ term in asymptotic
expansion (\ref{expansion_scheme}),
and appears to be the primary agent in the nonlinear saturation
of the MRI in the channel. It acts anisotropically so as
to modify the shear profile and results in a
non-diagonal stress component (relevant for radial angular momentum transport),
which behaves like $\sim 1/{\cal R}$.
\par
These results should be compared to experiments and their numerical
simulations.  Thus far no
experimental results for this geometry have been
reported but corresponding numerical simulations
do exist \cite{jeremy}.  Our results on the scaling of
transport (decreasing with increasing ${\cal R}$ and
independent of ${\cal R}_{\rm m}$ for
${\cal P}_{\rm m} \ll 1$) are qualitatively consistent
with these numerical simulations but more
careful comparisons are needed.
In addition,
issues of the effect of numerical resolution upon
the resulting dynamics
must be considered, as is done in the
turbulent dynamo problem  \cite{cata,brand2}, for example.
Our analysis is complementary to that of
Knobloch and Julien \cite{kj} who have recently reported
the results of an asymptotic MRI analysis
for a developed state far from marginality.
\par
The trends predicted by this simplified model
are not qualitatively altered by the appearance of boundary layers
which arise
when no-slip, perfectly conducting boundary conditions are
enforced
in the limit ${\cal P}_{\rm m}\ll 1$.  Calculations
with these more realistic boundary conditions
lead to amplitude equations, whose coefficients can not be, however, expressed
analytically in any convenient way. A full presentation of such a calculation will appear
in a forthcoming work.
\par
Further analytical investigations of the nonlinear MRI
will contribute
toward assembling a hierarchical understanding of this important instability,
from laboratory experiments to numerical simulations to
astrophysical disks.\par
We thank Jeremy Goodman for his critically constructive comments on a previous
version of this work and Michael Mond for insightful discussions.


\begin{thebibliography}{}
\bibitem{SSB}
    The SB approximation can be developed in a systematic way, using
    asymptotic scaling arguments \cite{we}. One of its limits
    is the ``small" shearing box (SSB) and in it the background fluid has constant
    density and the flow is incompressible (in addition to the canonical
    SB assumptions.)
\end{thebibliography}
\end{document}